\documentclass[12pt]{article}
\usepackage{epsfig}
  
\def\be{\begin{equation}}
\def\ee{\end{equation}}
\def\ben{\begin{displaymath}}
\def\een{\end{displaymath}}
\def\ba{\begin{array}{c}}
\def\bal{\begin{array}{l}}
\def\ea{\end{array}}

\begin{document}

\begin{center}

{\Large \bf Quantum toboggans: models exhibiting a multisheeted
${\cal PT}-$symmetry}

\vspace{2cm}

 {\bf Miloslav Znojil
}

NPI AS CR, 250 68 \v{R}e\v{z},

Czech Republic

{\it email: znojil@ujf.cas.cz}

\end{center}

\vspace{2cm}

\section*{Abstract}

A generalization of the concept of ${\cal PT}-$symmetric
Hamiltonians $H=p^2+V(x)$ is presented. For the usual analytic
potentials $V(x)$ (with singularities) and for the recently widely
accepted ``${\cal PT}-$symmetric" asymptotic boundary conditions
for wave functions $\psi(x)$ (selected inside a pair of complex
wedges generalizing the usual $x \to \pm \infty$ asymptotics),
non-equivalent quantum toboggans are defined as integrated along
topologically different paths ${\cal C}$ of coordinates $x\in
C\!\!\!\!l\ \ $. %Both the bound states and scattering solutions
%are discussed.

\section{Introduction}

Among non-Hermitian Hamiltonians $H \neq H^\dagger$ defined, say,
in a suitable ``auxiliary" Hilbert space ${\cal H}^{(aux)}$, a
privileged subfamily is formed by their $\eta-$pseudohermitian
special cases which are such that $ H^\dagger= \eta \,H
\,\eta^{-1}$ in terms of a suitable operator $\eta \neq I$.
Whenever $\eta={\cal P}$ happens to coincide with the operator of
parity, we arrive at the increasingly popular \cite{Carl} ${\cal
PT}-$symmetric quantum models~$H$.

In general, {\em all} of the above-mentioned Hamiltonians $H$ can
happen to possess a real and discrete spectrum of energies $E_0
\leq E_1 \leq \ldots$. In such a case we may always imagine that
while our $H \neq H^\dagger$ (acting in ${\cal H}^{(aux)}$) is
just one of many possible operator representations of such a
spectrum of bound-state energies, there may exist a ``true"
physical Hilbert space ${\cal H}^{(phys)}$ and, in it, an
orthonormalized basis of kets $\{\,|n\rangle\,\}_{n=0, 1, \ldots}$
such that a certain operator $H^{(phys)}$ defined by its spectral
representation,
 \be
 H^{(phys)} =
 \sum_{n=0}^\infty\,|\,n\rangle \,E_n\,\langle n |
 \ee
can be declared the ``true", mathematically correct representant
of the Hamiltonian of our original quantum system. Indeed, the
latter operator is manifestly {\em self-adjoint} in ${\cal
H}^{(phys)}$ and all the postulates of Quantum Mechanics are
satisfied.

There exist multiple practical applications of such an idea of the
use of two different Hilbert spaces in parallel (cf., e.g., the
most recent and up-to-date review \cite{Carl}). One of their most
characteristic shared features is that while $H^{(phys)}$ is a
technically very complicated operator in  ${\cal H}^{(phys)}$, its
``equivalent" representation  ${H}^{}$ in  ${\cal H}^{(aux)}$ is,
by assumption,  ``very simple", in comparison at
least~\cite{Geyer}.

In the latter sentence, we used the quotation marks on purpose:
``very simple" need not mean trivial. Even worse, the
``equivalence" of the mutual mapping between  ${H}^{}$ and
${H}^{(phys)}$ is particularly conventional a concept. Indeed, it
is obvious that in the very second paragraph of our present text
the readers may have noticed that our choice of the basis
$\{\,|n\rangle\,\}_{n=0, 1, \ldots}$ was {\em entirely} arbitrary.

Of course, a responsible attitude towards the latter ambiguity
problem unifies various applications of the formalism which may be
sampled as ranging from nuclear physics \cite{Geyer} and field
theory \cite{Carl} far beyond the territory of quantum theories,
involving even the terrains as distant as random matrices
\cite{Joshua}, cosmology \cite{Kamen} or classical optics
\cite{Berry}, electrodynamics \cite{aliben} and
magnetohydrodynamics \cite{Use}. In each of these contexts, the
ambiguity problem finds its specific resolution. In particular, in
the narrower domain of quantum theories themselves, various
additional conditions are usually imposed, relying on various
persuasive phenomenological arguments as summarized, in compact
form, in refs.~\cite{Geyer,Ali} (for Quantum Mechanics) and
\cite{Carl} (mainly in the context of field theories).

Once a suitable map between ${\cal H}^{(aux)}$ and ${\cal
H}^{(phys)}$ has been established (this is not to be discussed
here), the most persuasive distinction between these spaces can be
seen in the prohibitively complicated form of the operator
${H}^{(phys)}$ when compared with ${H}^{}$. Typically, in the
specific, ${\cal PT}-$symmetric quantum models~$H= {\cal
P}\,H^\dagger\,{\cal P}$, the manifestly self-adjoint
${H}^{(phys)}$ is even fairly difficult to define. In contrast,
${H}^{}$ is often selected as an ordinary differential operator,
easily and efficiently tractable by many standard mathematical
techniques. {\it Pars pro toto}, we recommend the readers to have
a look at the ``first nontrivial", exactly solvable ${\cal
PT}-$symmetric harmonic oscillator~\cite{ptho} for illustration.

On the latter, ${\cal PT}-$symmetric background there is still a
lot of space for the study of some ``slightly" more complicated
though still mathematically tractable Hamiltonians ${H}^{}$. In
this spirit we intend to pay attention here to the newly
introduced \cite{QT0} and developed \cite{QT1,QR2} family of the
models called ``quantum toboggans" (QT). Our brief review and
introduction to this subject will be divided in section \ref{L2}
(on the concept of quantum toboggans), section \ref{L3} (on QT
constructions), section \ref{L4} (on the QT models with more
branch points) %, section \ref{L5} (on the scattering solutions)
and summary.

\section{The concept of quantum toboggans \label{L2} }
%of  ${\cal PT}-$symmetric Quantum Mechanics

During a ``prehistory" of our theme, several complex potentials
$V(x)$ have been shown to generate real bound-state spectra, be it
an imaginary cubic $V(x) \sim {\rm i}x^3$ at $|x| \gg 1$
\cite{Caliceti} or a negative quartic  $V(x) \sim -x^4$ at $|x|
\gg 1$ \cite{BG}, both exhibiting ${\cal PT}-$symmetry (note that
${\cal P}$ changes parity in this context, $x \to -x$, while the
complex conjugation ${\cal T}$ mimics time reversal).

In some sense, the ``history" commenced in 1993 when Bender and
Turbiner published a letter where certain standard ordinary
differential Schr\"{o}dinger equations
 \be
 \left (
 -\frac{d^2}{dx^2} + V(x)
 \right )\,\psi(x)=E\,\psi(x)
 \label{SE}
 \ee
were declared physical {\em even when} complemented by certain
{anomalous, complexified} Dirichlet asymptotic boundary conditions
\cite{BT}. Indeed, once you assume a suitable form of analyticity
of $V(x)$ you may require, in a mathematically consistent manner,
that
 \be
 \psi \left (- \varrho \cdot e^{i\,\theta_{(left)}} \right ) =
 \psi \left (+ \varrho \cdot e^{i\,\theta_{(right)}} \right ) =  0
 \ \ \ \ \ \ \ \varrho \to +\infty\,
 \label{asybc}
 \ee
not only for the usual $\theta_{(left)}= \theta_{(left)}=0$ but
also at {\em any} non-vanishing angles $\theta_{(left)}\neq 0 \neq
\theta_{(left)}$.

\begin{figure}

  \begin{center}
\psfig{figure=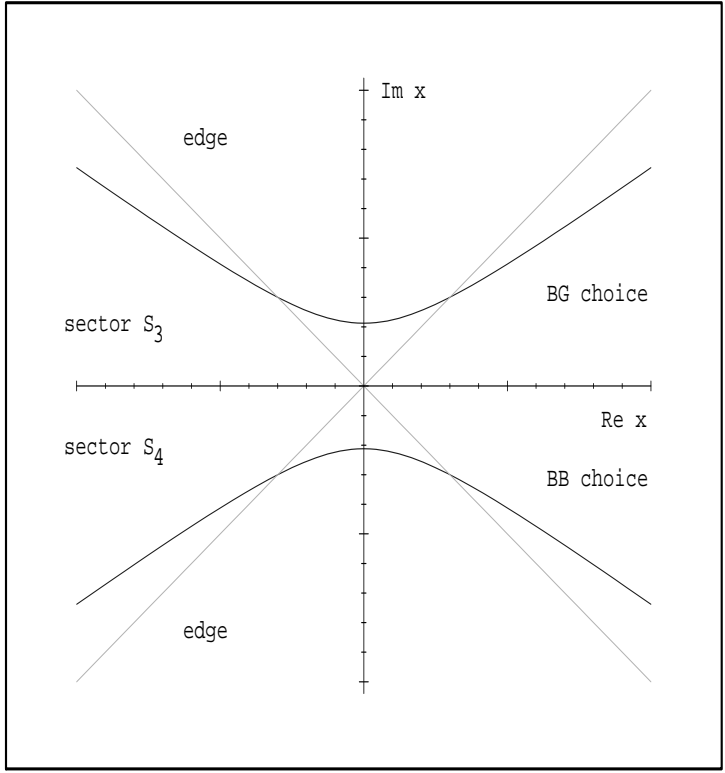,width=11.6cm, height=10.5cm,angle=-90}
\caption{Complex contours of coordinates
 (BG = choice made in ref.
 \cite{BG}, BB = choice made in ref. \cite{BBq}). } \label{map1}

\end{center}
\end{figure}

The next and decisive step towards possible explicit and
nontrivial applications of such a more or less trivial mathematics
in physics has been made by Bender and Boettcher \cite{BB} who
presented a persuasive numerical and semiclassical support for
their conjecture that in many similar cases the spectrum {\em can}
remain real and, hence, {observable}. More specifically, they
employed and recommended a special choice of the conditions
(\ref{asybc}) setting, in our present notation,
 \be
 \theta_{(left)}=-\theta_{(right)}\,.
 \label{ptasybc}
 \ee
They found out, purely empirically, that such a postulate seems to
offer very good chances that the spectrum remains real. In the
other words, in a way transferring certain numerical experience
\cite{DB} and known mathematical tricks (cf., e.g., the textbooks
\cite{xx} or the Buslaev's and Grecchi's paper \cite{BG}) into a
much more ambitious physical project. They recommended to work
with the ``curved", complex paths ${\cal C}$ of coordinates $x$
which remain left-right symmetric in the complex plane of $x$,
i.e.,  which are, in an obvious sense, ``${\cal PT}-$symmetric"
(cf. their sample given here in Figure~\ref{map1}).

%${\cal C}$ in $C\!\!\!\!l\ \ $

In a way fortunate for the subsequent quick development of the
subject \cite{xxx}, the reality of spectra in many ${\cal
PT}-$symmetric models has been fairly soon proved in an entirely
rigorous manner \cite{DDT}. The way has been opened for the birth
of quantum toboggans \cite{QT0}.

  \begin{figure}

  \begin{center}
\psfig{figure=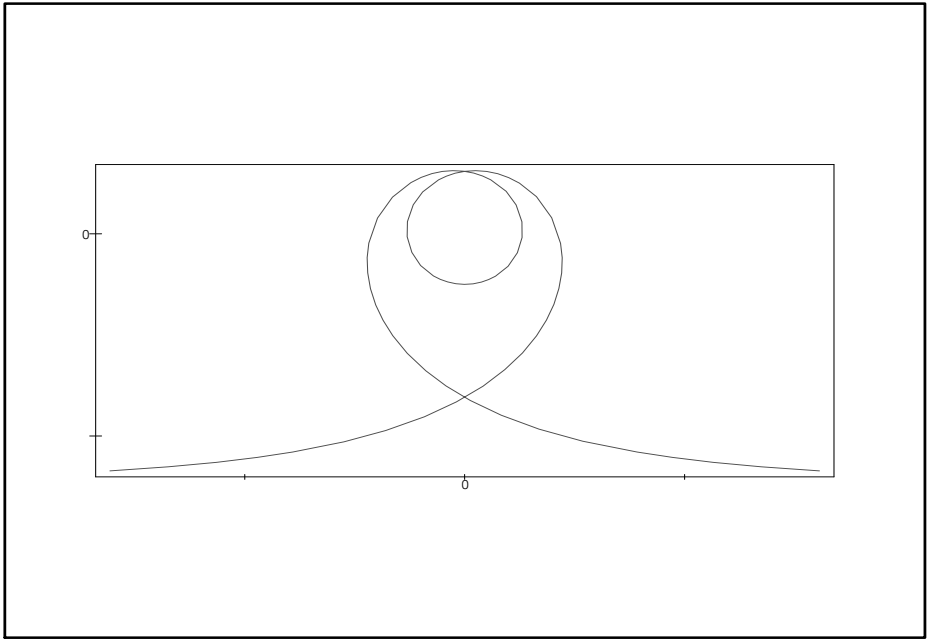,width=9.6cm, height=13.5cm,angle=-90}
\caption{ Tobogganic contour ${\cal C}^{(N)}$ with ${N}=2$. }
\label{map5}

\end{center}
\end{figure}

The mathematical essence of quantum toboggans (QT, \cite{QT1}) is
easy to explain briefly. On an intuitive level one immediately
sees that in the case of the presence of a branch point in
$\psi(x)$ (say, at $x^{(BP)}=0$), the ``standard" integration
paths as sampled in Figure \ref{map1} can be replaced, e.g., by
the ``tobogganic spiral" of Figure \ref{map5}. On a more abstract
level one simply has to recognize that the current (and
mathematically very strong) assumptions concerning the analytic
behaviour of $V(x)$  may be perceivably weakened without any real
changes in our understanding of the ${\cal PT}-$symmetry-related
problems, conjectures and constructions. {\it In nuce}, we may
admit that in the complex plane of $x$ our potentials may be
allowed to have singularities which are distributed in a
left-right symmetric manner in $C\!\!\!\!l\ \ $.

In all the cases with {\em more than one} branch point (cf.
\cite{QT1,QR2} or section \ref{L4} below), the discussion of the
non-equivalent QT paths would be technically complicated though
still very similar to the single-branch-point simplest case. For
this reason, let's now stay just in the simplest QT = QT1 case
with holomorphic $V(x)$ possessing the centrifugal-like pole $\sim
\ell(\ell+1)/x^2$ in the origin. Obviously, this model generates
also a branch point $x^{(BP)}=0$ in $\psi(x)$ generated by this
centrifugal-type term in the potential. In order to classify the
related non-equivalent integration paths ${\cal C}$ in
eq.~(\ref{SE}), one must check how many times they turn around the
centre before they start approaching their $ \varrho \gg 1$
asymptotics.

  \begin{figure}

  \begin{center}
\psfig{figure=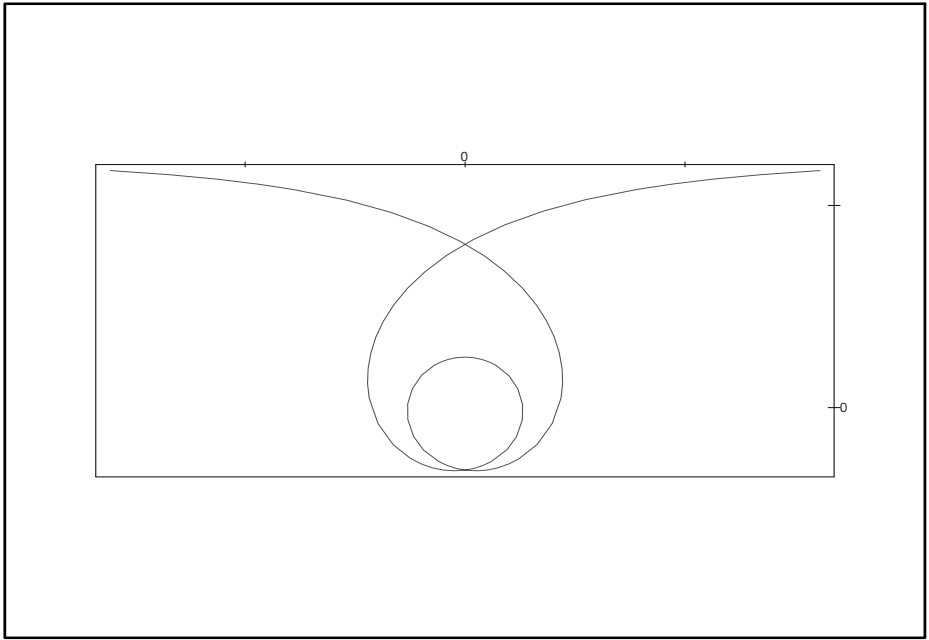,width=9.6cm, height=13.5cm,angle=90}
\caption{The complex conjugate version of the contour of Figure
\ref{map5}. } \label{imap5}

\end{center}
\end{figure}

This check is also sufficient. Indeed, as long as the coordinates
$x\in {\cal C}$ are complex and, by assumption, $0=x^{(BP)} \notin
{\cal C}$, we may just let the angles $\theta_{(left)}$ and
$\theta_{(right)}$ vary {\em beyond} the interval
$(-\pi/2,\pi/2)$. In particular, in the specific ${\cal
PT}-$symmetric cases as defined by eq.~(\ref{ptasybc}), we shall
accept a convenient convention concerning the (say,
counterclockwise) orientation of the winding of the curves ${\cal
C}$. Moreover, on the zeroth Riemann sheet of the total Riemann
surface pertaining to our QT bound states $\psi_(x)$ we shall fix
our QT curves ${\cal C}$ as crossing the imaginary axis below zero
and in the left-right direction so that, for example, the curves
running in the opposite direction (like the one  displayed in
Figure \ref{imap5}) will not be considered here as independent new
models.

\section{QT1 models with the single branch point  \label{L3} }
%={\cal C}^{(N)}

The overall winding number $N$ of ${\cal C}$ is highly relevant
for the specification of the boundary conditions~(\ref{asybc}). In
the presence of just single branch point $x^{(BP)}$, the knowledge
of the winding number $N$ in ${\cal C}={\cal C}^{(N)}$ specifies
the QT1 bound-state problem completely. We may visualize our QT1
spirals ${\cal C}^{(N)}$ as curves which are parametrized by an
angle $\phi \in (-\Phi, \Phi)$ with a positive radius
$\varrho=\varrho(\phi)$. In such a setting, the simplest
definition of the ${\cal PT}-$symmetry of the curve ${\cal
C}^{(N)}$ may be based on the symmetry requirement
$\varrho(\phi)=\varrho(-\phi)$ \cite{QT0}.

%{\bf II.  MODELS ON COMPLEX CONTOURS ${\cal C}^{(N)}$ }

For the most elementary ${\cal PT}-$symmetric harmonic-oscillator
Schr\"{o}dinger QT1 equation
 \ben
 \left (
 -\frac{d^2}{dx^2} + \frac{\ell(\ell+1)}{x^2} + x^2
 \right ) \,\psi (x)= E \,\psi (x)
   \label{HO}
 \een
${\cal PT}-$symmetric  path ${\cal C}^{(N)}$ N-times encircles
$x=0$. At the trivial $N=0$ this model may be defined along the
straight contour
 \ben
 {\cal C}^{(0)} = \left \{
 x\, \left | \, x = t-i\,\varepsilon ,\,
 t\in I\!\!R  \right .
 \right \}
 \label{BGC}
 \een
and one finds that at any $\alpha(\ell)=\ell+1/2 $ it generates
``twice as many" bound-state levels than its half-line predecessor
(cf. \cite{ptho}),
  \ben
 E=E_{n,\ell, \,\pm}^{}=4n+2 \pm 2\alpha(\ell)\,,
 \ \ \ \ \ n = 0, 1, \ldots\,.
 \een
Various anharmonic non-tobogganic generalizations of this solvable
model with various potentials such that ${\rm Re}\, V(x)=+{\rm
Re}\, V(-x)$ and ${\rm Im}\, V(x)=-{\rm Im}\, V(-x)$ cease to be
unique since even several non-tobogganic asymptotic boundary
conditions
 \ben
 \psi(\pm {\rm Re}\ L +i\,{\rm Im }\ L) = 0\,,
 \een
 \ben \ \
  \ \ \ \ \ \ |L| \gg 1\ \ \ \ {\rm or}
  \ \ \ \ |L| \to \infty\,.
 % \label{bcs}
  \een
can prove non-equivalent.

%
% \ben
% {\cal C}^{(N)} ={\cal D}_{(\varepsilon,N)}^{(PTSQM,\,tobogganic)}
% \een
%\newpage
%
%{\bf new quantum theory}:

\begin{figure}

  \begin{center}
\psfig{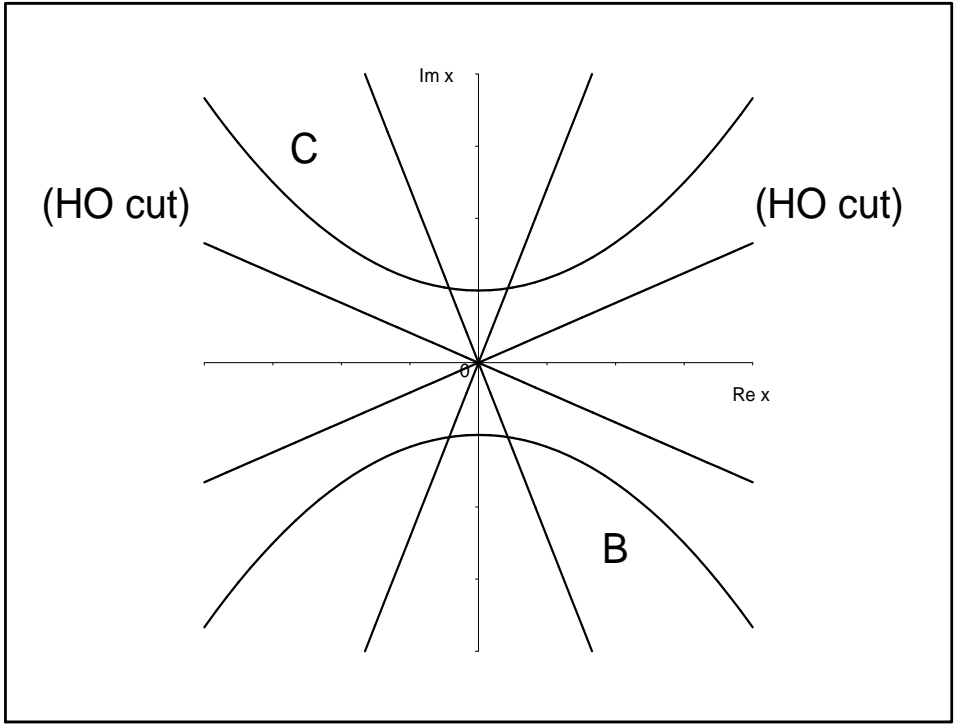}
\caption{Sextic oscillator as a map of a ${\cal PT}-$symmetric
harmonic-oscillator contour ${\cal C}^{(0)}$  (curve B) and of a
tobogganic harmonic oscillator contour ${\cal C}^{(1)}$ (curve C).
} \label{dmap1}

\end{center}
\end{figure}

In the tobogganic cases characterized by a nonvanishing
winding-number integer $N \neq 0$, the curves ${\cal C}={\cal
C}^{(N)}$ lie on a multisheeted Riemann surface. They can be
parametrized there by an angle $\varphi \in
 \left (
 -(N+1)\pi,\,N\pi^{}
 \right )$ as, e.g.,
 \ben
  {\cal C}^{(N)} =
 \left \{
 x = \varepsilon\,
 \varrho(\varphi,N)\,e^{i\,\varphi} \,,  \left .
  \varepsilon > 0 \right .
 \right \}\,,
 \label{BGCN}
 \een
\ben
  \ \varrho(\varphi,N)=\sqrt{1 + \tan^2 \frac{\varphi+\pi/2}
 {2N+1}}\,.
 \een
A deeper discussion of some features of quantum toboggans can be
facilitated by the quasi-exact solvability of the underlying
potential chosen, say, in the asymptotically decadic form
 \ben
 V(x) = x^{10} + {\rm asymptotically \ smaller\ terms}\,
 \label{asydec}
 \een
where \cite{Znojilqes}
 \ben
 \psi(x) = e^{-x^6/6+ {\rm asymptotically \ smaller\ terms}}\,.
  \een
We may reparametrize
 \ben
 \psi(x) = \exp
 \left [-\frac{1}{6}\varrho^6 \cos 6\varphi
 + \ldots
 \right ]\,
 \een
and see that there exist as many as five non-tobogganic versions
of this model, with angles in the eligible complex wedges
 \ben
 \Omega_{(first\ right)} =
 \left (
 -\frac{\pi}{2}
 +\frac{\pi}{12},
 -\frac{\pi}{2}
 +\frac{3\pi}{12}
 \right ),\ \ \ \
 \een
 \ben
 \Omega_{(first\ left)} =
 \left (
 -\frac{\pi}{2}
 -\frac{\pi}{12},
 -\frac{\pi}{2}
 -\frac{3\pi}{12}
 \right ),
 \een
 \ben
 \Omega_{(third\ right)} =
 \left (
 -\frac{\pi}{2}
 +\frac{5\pi}{12},
 -\frac{\pi}{2}
 +\frac{7\pi}{12}
 \right ),\ \ \  \ldots\
 \een
 \ben \ \
 \ldots\ \ \ \  \Omega_{(fifth\ left)} =
 \left (
 -\frac{\pi}{2}
 -\frac{9\pi}{12},
 -\frac{\pi}{2}
 -\frac{11\pi}{12}
 \right )\,.
 \een
It is worth noting that {\em all} of the models of this type can
be interpreted as equivalent to their non-tobogganic partners
confined by a {\em different} ``effective" potential. For this
purpose one can simply perform a  ${\cal PT}-$symmetric change of
variables in the ``initial" ${\cal PT}-$symmetric model
 \ben
 \left [
 -\frac{d^2}{dx^2} - (ix)^2 + \lambda\,W(ix)
  \right ]
 \,\psi
 (x)=
 E(\lambda)
 \,\psi
 (x)\,
 \een
where
 \ben
 \ \ \ \ \ \ W(ix)={\bf \Sigma}\,g_\beta(ix)^\beta\,
 \label{AHO}
 \een
and where one sets
 \ben
 ix = (iy)^\alpha\,, \ \ \ \ \ \ \
 \psi(x) = y^\varrho\,\varphi(y).
 \label{change}
 \een
Once we use the freedom in the choice of $\alpha>0$ we have
 \ben
 i \,dx = i^\alpha \alpha y^{\alpha-1}\,dy
 , \ \ \ \ \ \ \ \
  \frac{(iy)^{1-\alpha}}{\alpha}\,\frac{d}{dy}= \frac{d}{dx}\,.
  \een
This gives an equivalent, ``Sturmian" problem in an intermediate
differential equation form
 \ben
 y^{1-\alpha}
 \frac{d}{dy}
 y^{1-\alpha}
 \frac{d}{dy}\,y^\varrho\,\varphi(y)+
 \een
 \ben
 +i^{2\alpha} \alpha^2 \left [
 - (iy)^{2\alpha} +
 \lambda\,W[(iy)^{\alpha}]-
 \right .
 \een
 \ben
  \left . -E(\lambda)
  \right ]
 \,y^\varrho\,\varphi(y)=0\,.
 \een
Here, the first term
 \ben
y^{1-\alpha}
 \frac{d}{dy}
 y^{1-\alpha}
 \frac{d}{dy}\,y^{[(\alpha-1)/2]}\,\varphi(y)
 =
 \een
 \ben
 =y^{2+\varrho-2\alpha} \frac{d^2}{dy^2}\,\varphi(y)
 +\varrho(\varrho-\alpha)y^{\varrho-2\alpha}\,\varphi(y)\,,
 \een
``behaves" at the specific
 \ben \ \ \ \ \ \  \ \ \
 \varrho = \frac{\alpha-1}{2}\,
 \een
so that the new equation preserves the same Schr\"{o}dinger form:
 \ben
- \frac{d^2}{dy^2}\,\varphi(y)
 +\frac{\alpha^2-1}{4y^{2}}
 \,\varphi(y)+ \een \ben
 +(iy)^{2\alpha-2} \alpha^2 \left [
 - (iy)^{2\alpha} +
 \lambda\,W[(iy)^{\alpha}]\,\varphi(y)=
 \right .
 \een
 \ben
 =(iy)^{2\alpha-2} \alpha^2 \, E(\lambda)
  \,\varphi(y)\,.
 %\label{AHOnew}.
 \een
What was important is that the change of variables
 changed also the range of the angle in ${\cal C}^{(N)}$
so that by the choice of $\alpha$ one can diminish the winding
number $N$. In this manner, many polynomial potentials prove
interrelated. E.g., with $\alpha = 1/2$ we get the quadratic
oscillator $V_g(y)=
 -({iy})^{2} +
 i\,g_{1}\,{y}+
  g_{-1}\,({iy})^{-1}+
  g_{-2}\,({iy})^{-2}$
from the sextic oscillator
 \ben
- \frac{d^2}{dx^2}\,\varphi(x)
 +\frac{\ell(
 \ell+1)}{{x}^{2}}
 \,\varphi({x})
 +V_f(x)
 % \left [
% {x}^{4q+2} +
% g_{4q}\,{x}^{4q}+
% \ldots
%  +
% g_2\,{x}^{2}
%  \right ]
 \,\varphi({x})=
 E\,\varphi({x})
 \,,
\label{SEget}
 \een
 \ben
 V_f(x)=
 {x}^{6} +
 f_{4}\,{x}^{4}+
  f_2\,{x}^{2}+
 f_{-2}\,{x}^{-2}, \ \ \ \ \
 \een %\ben
% V_g(y)=
% -({iy})^{2} +
% i\,g_{1}\,{y}+
%  g_{-1}\,({iy})^{-1}+
%  g_{-2}\,({iy})^{-2}.
%  \een
%  \ben
%  \ \ \ \ \
% V_h(y)=
% -({iy})^{2/3} +
% h_{-2/3}\,({iy})^{-2/3}+
%  h_{-4/3}\,({iy})^{-4/3}+
%  h_{-2}\,({iy})^{-2}
% \een
etc (cf. Figure \ref{dmap1}).

In the conclusion of this section let us emphasize that the ${\cal
PT}-$symmetry in the presence of the single branch point can be
based on the introduction of the two different parity-like
operators ${\cal P}^{(\pm)}: x \to x \cdot \exp (\pm i\pi)$ as
well as of the two eligible rotation-type innovations ${\cal
T}^{(\pm)}$ of the time reversal.

%
%same for ${\cal P}^{(\pm)}{\cal T}^{(\pm)}$ and
% \ben
%  \left ({\cal C}^{(N)}\right )^\dagger
%  ={\cal D}_{(\varepsilon',N)}^{(PTSQM,\,tobogganic)}\,,
%  \ \ \ \ \ \
% \varepsilon'=\varepsilon\cdot e^{\pm i\pi}
%  \,.
% \een

%
%\section*{Bound states  \label{IIId}}
% \ben
% H_{({\cal PT})}\,\psi(x)=E\,\psi(x)
% \een
%with Dirichlet inside the wedges,
% \ben
% \psi \left ( \varrho \cdot e^{i\,\theta} \right ) = 0,
% \ \ \ \ \ \ \ \varrho \gg 1
%  \label{bc}\ \ \ \ \ \
% \theta + k_{i,f}\,\pi \in
%  \left (
% -\frac{\pi}{4},\frac{\pi}{4}
%  \right )\,
%  \een
%spectra = real in unbroken cases.
%

\section{QT models with more branch points  \label{L4} }
%{\bf V. TOBOGGANS IN POTENTIALS WITH MORE SPIKES \label{IV.1}}

Once you study the bound-state problem
\begin{equation}
-\frac{\hbar ^2}{2m}\,\frac{d^2}{dx^2}\,\psi _n(x)+V(x)\,\psi
_n(x)=E_n\,\psi _n(x)\,,  \label{SEd}
\end{equation}
you may set $\hbar=2m=1$ and choose, say, the potential with two
second-order poles,
 \ben
 V(x) = V_{regular}(x)+ \frac{G}{(x-1)^2} + \frac{G^*}{(x+1)^2}\,.
 \een
In this way you arrive at the wave functions
\begin{equation}
\psi ^{(general)}(x)=c_{+}\psi _{(+)}^{(special)}(x)+c_{-}\psi
_{(-)}^{(special)}(x)\,  \label{wavefs}
\end{equation}
with the two (in general, complex) branch points at, say,
$x=x^{(BP)} = \pm 1$ \cite{Anjaja}. Thus, the Riemann surface
$\mathcal{R}$ of $\psi (x)$ becomes composed of {\em many} sheets
${\mathcal R}_k$.

\subsection{The harmonic-oscillator example}

Without boundary conditions our differential Schr\"{o}dinger
equation comprises many eigenvalue problems at once \cite{BT}. In
the QT = QT2 models with two branch points they are all in a
one-to-one correspondence with our selection of the QT paths
$x^{(\varrho)}(s)$. In a way  introduced in ref. \cite{QR2}, they
may be classified by certain ``winding" or ``knotting" descriptors
$\varrho $. These $x^{(\varrho)}(s)$ connect their asymptotes
while passing through a compact domain of $x$ which contains all
the singularities of $V(x)$.

In the most transparent non-tobogganic  case one can stay on the
single Riemann sheet and, for illustrative purposes, pick up the
harmonic oscillator with $\psi _n^{(\pm |\alpha )|}(x)=x^{1/2\pm
|\alpha |}\,\exp (+x^2/2)\times \mathrm{a\,polynomial}$,
considered as integrated along U-shaped paths
\begin{equation}
y^{(U)}(s)=x^{(U)}(s)+\mathrm{i}\,\varepsilon =\,\left\{
\begin{array}{ll}
|\,s|\,e^{-11\,\mathrm{i}\,\pi /8}\exp [\mathrm{i}\,\xi (s)],\, &
\ \ s\ll -1\,, \\ |\,s|\,e^{3\mathrm{i}\,\pi /8}\exp
[\mathrm{i}\,\xi (s)], & \ \ s\gg 1\,
\end{array}
\right.   \label{Supshift}
\end{equation}
on which $x(s)=y(s)-\mathrm{i}\,\varepsilon $ , $\xi (s)\in (-\pi
/8,\pi /8)$ and
\begin{equation}
\lim_{s\to \pm ,\infty }\,\psi \left[ \,x^{(U)}(s)\right] =0\,.
\label{SEHOanobc}
\end{equation}
In the alternative, tobogganic cases with $N=1$, $\xi \in (-\pi
/8,\pi /8)$ and  $\delta >0$ in
\begin{equation}
x^{(N=1)}(s)=y^{(N=1)}(s)-\mathrm{i}\,\varepsilon =\,\left\{
\begin{array}{ll}
|\,s-\eta |\,e^{-13\,\mathrm{i}\,\pi /8}\exp [\mathrm{i}\,\xi (s)]+\mathrm{i}%
\,\delta ,\, & \ \ s\ll -1\,, \\
|\,s-\eta |\,e^{5\mathrm{i}\,\pi /8}\exp [\mathrm{i}\,\xi (s)]+\mathrm{i}%
\,\delta , & \ \ s\gg 1\,
\end{array}
\right.   \label{Uptobshift}
\end{equation}
we may consider the  paths  encircling two branch points  by
winding

\begin{itemize}
\item  counterclockwise around $x_{(-)}^{(BP)}$ (to be marked by a
letter $L$),

\item  counterclockwise around $x_{(+)}^{(BP)}$ (letter $R$),

\item  clockwise around $x_{(-)}^{(BP)}$ ($Q=L^{-1}$),

\item  clockwise around $x_{(+)}^{(BP)}$ ($P=R^{-1}$).

\end{itemize}

 \noindent
In this way, a four-letter alphabet can be used to label all paths
${x}=x^{(\varrho )}(s)$ by words $\varrho $ of length $ 2N$  and
of a concatenated form $\varrho =\Omega \bigcup \Omega ^T$ which
is due to the underlying $\mathcal{PT}-$symmetry $L\leftrightarrow
R$.

Thus, besides the symbol $\varrho =\emptyset$ for the
non-tobogganic case one has  four possibilities at $N=1$, viz.,
\[
\Omega \in \left\{ L\,,L^{-1}\,,R\,,R^{-1}\right\} \,,\ \ \ \ \
N=1\,,
\]
\[
\varrho \in \left\{ LR\,,L^{-1}R^{-1}\,,RL\,,R^{-1}L^{-1}\right\}
\,,\ \ \ \ N=1\,,
\]
or the following dozen cases at $N=2,$
\[
\Omega \in \left\{ LL,LR,RL,RR,L^{-1}R,R^{-1}L,LR^{-1}\,, \right .
\]
\[
\left . R\,L^{-1}
\,,L^{-1}L^{-1},L^{-1}R^{-1},R^{-1}L^{-1},R^{-1}R^{-1}\right\}
\]
(with four items $LL^{-1}\,,L^{-1}L\,,RR^{-1}$, $R^{-1}R$ not
allowed among the $4^2=16$ eligible ones), etc \cite{QR2}.

%
%\fbox{at $N=3,$ } total number = 36:
%
% cross  28 out of
%$4^3=64$ words,
%
%$\Omega ^{(NA)}$ $=\Omega ^{(NAL)}\bigcup \Omega ^{(NAR)}$ (prev.
%$L,R$)
%
%$\Omega ^{(NAL)}=\Omega ^{(NAL3)}\bigcup \Omega ^{(NAL2)}$
%
%one or two inversions in  $\Omega ^{(NAL3)}$ (six words),
%
%in $\Omega ^{(NAL2)}$ add  $R$ or $R^{-1}$ (eight words).
%
%
%\fbox{at $N=4$ } we have  $256-76-40=140$:
%
%14 elements in $\Omega ^{(NAL4)}$
%
%24 elements in $\Omega ^{(NAL3)}$,
% $L\leftrightarrow R$
%
%
%%set $\Omega ^{(NAL2)}$ =
%
%$\Omega ^{(NAL21)}$ (single inversion, 16 elements),
%
%$\Omega ^{(NAL22)}$ (two inversions, 8 elements)
%
%$\Omega ^{(NAL23)}$ (three inversions, 16 elements).

%
%
%  \begin{figure}
%
%  \begin{center}
%\psfig{figure=fig55d.ps,width=9.6cm, height=12.6cm,angle=-90}
%\caption{Both the HO-cut lines move upwards, contour C becomes
%tobogganic} \label{ddmap1}
%
%\end{center}
%\end{figure}

\subsection{The rectification of the QT2 contours
at $\varrho =\varrho _0$}

In the presence of a single branch point we set
 \ben
 \mathrm{i}\,x=(\mathrm{i}\,z)^2\,,\ \psi _n(x)=\sqrt{z}\,
 \varphi_n(z)\, \label{rectifi1}
 \een
and remind the readers about the \emph{strict equivalence} of the
QT1 harmonic oscillator to its {manifestly
$\mathcal{PT}-$symmetric} sextic-oscillator partner
 \ben
 \left(-\frac{d^2}{dz^2}+4z^6+4E_nz^2+\frac{4\alpha^2-1/4}{z^2}
 \right)\varphi(z)=0
 \een
defined along a \emph{manifestly non-tobogganic} path,
\begin{equation}
 {\cal C}=
\left\{
\begin{array}{ll}
\sqrt{|\,s-\eta |}\,e^{-9\,\mathrm{i}\,\pi /16}\exp
[\mathrm{i}\,\xi (s)/2]+ \mathcal{O}(\delta /\eta ),\, & \ \ s\ll
-1\,, \\ \sqrt{|\,s-\eta |}\,e^{\mathrm{i}\,\pi /16}\exp
[\mathrm{i}\,\xi (s)/2]+ \mathcal{O}(\delta /\eta ), & \ \ s\gg
1\,
\end{array}
\right. .  \label{Uptoshit}
\end{equation}
In the presence of the {\em pair} of the branch points, say,
$x^{(BP)}=\pm 1$, the simplest changes of variables can be
employed again. To the Schr\"{o}dinger equation
 \ben
 %\begin{equation}
\left[ -\frac{d^2}{dx^2}+\frac{\ell (\ell +1)}{(x-1)^2}+\frac{\ell
(\ell +1)
}{(x+1)^2}+V(ix)\right] \,\psi (x)=
\een
\ben
=E\,\psi (x)\,.  %\label{inio}
%\end{equation} zero-energy and state-dependent
 \een
this often enables us to assign the rectified partner
 \ben
%\begin{equation}
\left[ -\frac{d^2}{dz^2}+U_{eff}(\mathrm{i}\,z)\right] \,\varphi
(z)=0\, \label{fino}
%\end{equation}
 \een
where
\[
U_{eff}(\mathrm{i}\,z)=U(\mathrm{i}\,z)+\frac{\mu (\mu
+1)}{(z-1)^2}+\frac{ \mu (\mu +1)}{(z+1)^2}\ \equiv \
\]
\[
\equiv \  U(\mathrm{i}\,z)+2\,\frac{\mu (\mu +1)[1-(
\mathrm{i}\,z)^2]}{\left[ 1+(\mathrm{i}\,z)^2\right] ^2}\,.
\]

\subsection*{Proof}

Using an implicit rectification formula
\[
1+(ix)^2=\left[ 1+(iz)^2\right] ^\kappa \,,\ \ \ \ \ \ \ \ \kappa
>1\,
\]
we reveal that $z=-\mathrm{i}\,\varrho $ gets mapped upon itself.
Hence, one can recommend the use of the explicit rectification
formula $ x=-\mathrm{i}\,\sqrt{(1-z^2)^\kappa -1}\,$. As a result,
certain effective non-tobogganic  potentials are obtained. Their
construction is routine since
 \begin{eqnarray*}
 \frac d{d\,x} &=&\beta (z)\,\frac d{d\,z}\,,\
 \ \ \ \ \ \beta (z)=-\mathrm{i}%
\,\frac{\sqrt{(1-z^2)^\kappa -1}}{\kappa \,z\,(1-z^2)^{\kappa
-1}}\,. \\ &&
\end{eqnarray*}
Thus, we may set $\psi (x)=\chi (z)\,\varphi (z)$  with $\chi
(z)=const\,/\sqrt{\beta (z)}$ \cite{Liouville} and get
$V_{eff}(\mathrm{i}\,x)=V(\mathrm{i}\,x)+2\,\ell (\ell
+1)[1-(\mathrm{i}\,x)^2]/[1+(\mathrm{i}\,x)^2]^2$ in
\[
\left( -\beta (z)\,\frac d{dz}\,\beta (z)\,\frac
d{dz}+V_{eff}[ix(z)]-E\right) \,\chi (z)\,\varphi (z)=0\,.
\]
or
 \ben
%\begin{equation}
U_{eff}(\mathrm{i}\,z)=\frac{V_{eff}[\mathrm{i}\,x(z)]-E_n}{\beta ^2(z)}+%
\frac{\beta ^{\prime \prime }(z)}{2\,\beta (z)}-\frac{[\beta ^{\prime }(z)]^2%
}{4\,\beta ^2(z)}  \label{last}
%\end{equation}
 \een
in the standard Schr\"{o}dinger equation.
 QED.

\begin{figure}

  \begin{center}
\psfig{figure=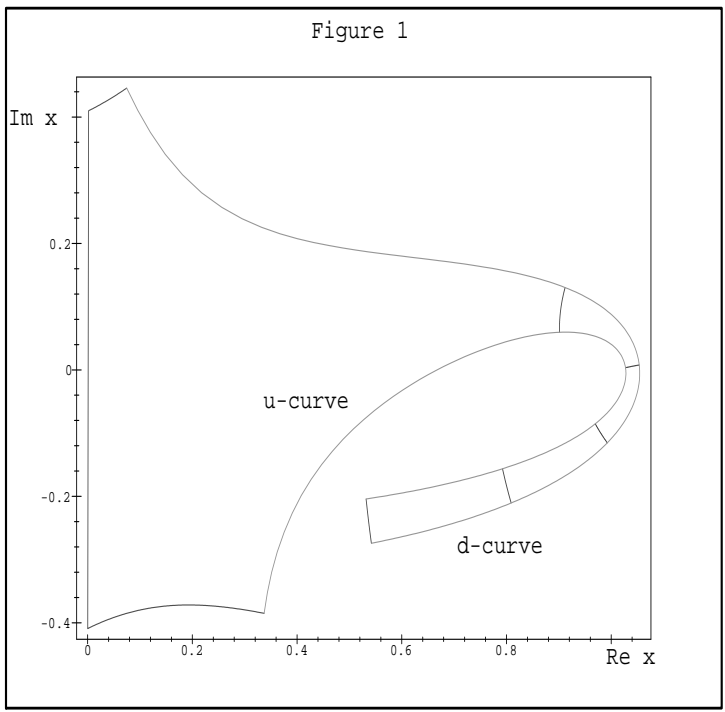,width=10.6cm, height=10.5cm,angle=-90}
\caption{Two bitoboggans ($\kappa =2.4$, $s\in (0.4,1.4)$). }
\label{smap1}

\end{center}
\end{figure}

Graphically, it is interesting to reconstruct the shapes of the
tobogganic pull-backs (cf. Figure \ref{smap1} for ilustration).
For this purpose, in the vicinity of the negative imaginary axis
we may consider the mapping
\[
z=-\mathrm{i}\,{r}\,e^{\mathrm{i}\,\theta }\ \longrightarrow
 \ \ x=-\mathrm{i%
}\,\left[ \left( 1+{r}^2\,e^{2\,\mathrm{i}\,\theta }\right)
^\kappa -1\right] ^{1/2}\,.
\]
At the small radii ${r}$ it degenerates to the mere multiplication
by a constant $\sqrt{\kappa}$. At the larger radii we arrive at a
more complicated knot-like shapes of $x^{\varrho _0}(s)$,
tractable easily by  computer graphics, via their definition as a
pullback of the straight-line $z(s)=s-\mathrm{i}\,\varepsilon $.
In this approach the winding number $N$ proves fairly sensitive to
the value of shift $\varepsilon $. In Figure~\ref{smap1} we  see
the clear acceleration of the winding after transition from
$\varepsilon =\varepsilon _u=0.15$ to $\varepsilon =\varepsilon
_d=0.20$.

%In
%Figure~3, the choice of $\kappa =5$ makes our spirals to turn
%twice during a small change of their parameter $s\in
%(0.55,1.285)$.

\section{Summary}

In the context of mathematics and in a way paralleling the birth
of interest in ${\cal PT}$ symmetry, the concept of quantum
toboggans could in fact be also understood as almost trivial in
mathematics. In physics, the first hints for its introduction
resulted from several independent mathematical sources. The first
one has been the above-cited Buslaev's and Grecchi's paper
\cite{BG} where a {\em formal necessity} of a branch point
involved {\em all} their analytic bound-state wave functions
$\psi_n(x)$. A clearer understanding of this hint (reflecting the
necessity of an effective kinematical centrifugal force in more
dimensions) came a few years later when attention has been turned
to an ``unperturbed" harmonic-oscillator special case of their
model \cite{ptho}. In parallel, virtually the same branch points
in wave functions resurfaced during the proofs of the reality of
the energies \cite{DDT}, during the studies of the quasi-exact
solvability of certain ${\cal PT}-$symmetric models
\cite{Znojilqes} and after a supersymmetrization of certain ${\cal
PT}-$symmetric Hamiltonians~\cite{Anjaja}.

Let us summarize: what should be remembered in the context of
mathematics is the new use of the changes of variables in
Schr\"{o}dinger equations. This can rectify the QT paths of
coordinates and may also lead to some  new and nonstandard {\em
feasible} calculations.

In the parallel phenomenological model-building context, the
perspective of new physics may be expected to be derived from the
prospective use of the tobogganic paths ${\cal C}^{(N)}$. This
could throw new light not only on ``innovated" bound states  (of a
``topological" origin) but also on the very unusual
scattering-type states \cite{QR2}.

\section*{Figure captions}

\subsection*{Figure \ref{map1}. Complex contours of coordinates
 (BG = choice made in ref.
 \cite{BG}, BB = choice made in ref. \cite{BBq}). }

\subsection*{Figure \ref{map5}. Tobogganic contour ${\cal C}^{(N)}$ with
${N}=2$.
}

\subsection*{Figure \ref{imap5}. The complex conjugate version of
the contour of Figure \ref{map5}. }

\subsection*{Figure \ref{dmap1}. Sextic oscillator as a map of a
${\cal PT}-$symmetric
harmonic-oscillator contour ${\cal C}^{(0)}$  (curve B) and of a
tobogganic harmonic oscillator contour ${\cal C}^{(1)}$ (curve C).
}

\subsection*{Figure \ref{smap1}. Two bitoboggans ($\kappa =2.4$, $s\in
(0.4,1.4)$).
}

%
%\subsection*{Figure \ref{smadp1}. Two bitoboggans ($\kappa =3$, $s\in
%(0.4,1.4)$).
%}
%
%\subsection*{Figure \ref{smad1}. Two bitoboggans ($\kappa =5$, $s\in
%(0.4,1.4)$).
%}

\vspace{5mm}

\section*{Acknowledgement}

Work supported by the GA\v{C}R grant Nr. 202/07/1307, by the
M\v{S}MT ``Doppler Institute" project Nr. LC06002 and by the
Institutional Research Plan AV0Z10480505.


\begin{thebibliography}{oo}

\bibitem{Carl}
C. M. Bender, Reports on Progress in Physics 70 (2007) 947.
%-1018 (2007) in print
%(hep-th/0703096);
%cf. also

\bibitem{Geyer}  F. G. Scholtz, H. B. Geyer and F. J. W. Hahne, Ann. Phys.
(NY) 213 (1992) 74.

\bibitem{Joshua}
A. Jarosz and M. A. Nowak, J. Phys. A: Math. Gen. 39 (2006) 10087.

\bibitem{Kamen}
A. A. Andrianov, F. Cannata and A. Y. Kamenshchik, J. Phys. A:
Math. Gen. 39 (2006) 9975.

\bibitem{Berry}
M. V. Berry, Czech. J. Phys. 54 (2004) 1039.

\bibitem{aliben}
A. Mostafazadeh, submitted.

\bibitem{Use}
O. Kirillov and U. G\"{u}nther, J. Phys. A: Math. Gen. 39 (2006)
10057.
%
%Krein space related perturbation theory for MHD alpha-2-dynamos

\bibitem{Ali}
A. Mostafazadeh, J. Math. Phys. 43 (2002) 205 and 2814 and 3944.

\bibitem{ptho}
M. Znojil, Phys. Lett. A 259 (1999) 220.

\bibitem{QT0}
M. Znojil,
  Phys. Lett. A 342 (2005) 36.
%  - 47
% (quant-ph/0502041):
%{\it PT-symmetric quantum toboggans

\bibitem{QT1}
M. Znojil,%  quant-ph/0606166
  J. Phys. A: Math. Gen. 39 (2006) 13325.
  % -  13336

\bibitem{QR2}
M. Znojil, Phys. Lett. A, to appear.

\bibitem{Caliceti}
E. Caliceti, S. Graffi and M. Maioli, Commun. Math. Phys. 75
(1980) 51.

\bibitem{BG}
V. Buslaev and V. Grecchi, J. Phys. A: Math. Gen. 26 (1993) 5541.
%Buslaev V and Grecchi V 1993 J. Phys. A: Math. Gen. 26 5541
%N. Hatano, D. R. Nelson, Phys. Rev. Lett. 77, 570 (1996);

\bibitem{BT}
C. M. Bender and A. Turbiner, Phys. Lett. A 173 (1993) 442.


\bibitem{BB}
C. M. Bender and S. Boettcher, Phys. Rev. Lett. { 80} (1998) 5243

\bibitem{DB}
D. Bessis, private communication (1992).

\bibitem{xx}
Y. Sibuya, Global Theory of Second Order Linear Differential
Equation with Polynomial Coefficient, North Holland, Amsterdam,
1975.

\bibitem{BBq}
C. M. Bender and S. Boettcher,  J. Phys. A: Math. Gen. 31 (1998)
L273.


\bibitem{xxx}
M. Znojil, Ed., Pseudo-Hermitian Hamiltonians in Quantum Physics
(IOP Prague, 2006, special issue Nr. 9 of Czech. J. Phys. (vol.
56)).

\bibitem{DDT}
P. Dorey, C. Dunning and R. Tateo, J. Phys. A: Math. Gen. 34
(2001) 5679;

K. C. Shin, Commun. Math. Phys. 229 (2002) 543.

\bibitem{Znojilqes}
M. Znojil, J. Phys. A: Math. Gen. 33 (2000) 6825.

\bibitem{Anjaja}
A. Sinha and P. Roy, Czech. J. Phys. 54 (2004) 129.

\bibitem{Liouville}
L. Liouville, J. Math. Pures Appl. 1 (1837) 16.

\end{thebibliography}
\end{document}